\documentclass[conference]{IEEEtran}
\IEEEoverridecommandlockouts
\usepackage{cite}
\usepackage{amsmath,amssymb,amsfonts}
\usepackage{algorithmic}
\usepackage{graphicx}
\usepackage{textcomp}
\usepackage{xcolor}
\def\BibTeX{{\rm B\kern-.05em{\sc i\kern-.025em b}\kern-.08em
    T\kern-.1667em\lower.7ex\hbox{E}\kern-.125emX}}

\begin{document}

\makeatletter
    \newcommand{\linebreakand}{%
      \end{@IEEEauthorhalign}
      \hfill\mbox{}\par
      \mbox{}\hfill\begin{@IEEEauthorhalign}
    }
\makeatother

\title{WIP: Exploring Student-Created AI Agents and Instructional Videos to Foster Engagement in Computing Education
}

\author{
\IEEEauthorblockN{Lucas G. Zangari}
\IEEEauthorblockA{\textit{School of Computing Instruction} \\
\textit{Georgia Institute of Technology}\\
Atlanta, GA, USA \\
lzangari3@gatech.edu}

\and

\IEEEauthorblockN{Olufisayo Omojokun}
\IEEEauthorblockA{\textit{School of Computing Instruction} \\
\textit{Georgia Institute of Technology}\\
Atlanta, GA, USA \\
omojokun@cc.gatech.edu}

\and

\IEEEauthorblockN{Pedro Guillermo Feij\'oo-Garc\'ia\textsuperscript{\textsection}}
\IEEEauthorblockA{\textit{School of Computing Instruction} \\
\textit{Georgia Institute of Technology}\\
Atlanta, GA, USA \\
pfeijoogarcia@gatech.edu}

}

\maketitle
\begingroup\renewcommand\thefootnote{\textsection}
\footnotetext{Corresponding author}
\endgroup
\begingroup\renewcommand\thefootnote{*}
\endgroup

\begin{abstract}
This work-in-progress (WIP) research-to-practice study examines the use of student-created artifacts to promote active learning in a large undergraduate Software Design and Engineering (SWE) course. Active learning strategies have been shown to significantly enhance student engagement, comprehension, and motivation, particularly in computing education. In this work, we explore the use of student-created artifacts to foster student engagement in higher-level Computer Science (CS) courses. Specifically, we examine the experiences of undergraduate and graduate CS students who participated in the creation of two types of artifacts: instructional video tutorials and conversational AI agents. The goal being to understand how designing these artifacts could support students in applying and learning course content. Students raised valuable insights regarding both experiences. Concerning the creation of instructional videos, they appreciated the interactivity brought through multimedia to their learning experiences, although some found the strategy to be time-consuming. Similarly, when designing conversational AI agents, students enjoyed the novelty and practical application of the task, while some encountered challenges related to the complexity of tools such as \textit{Google Dialogflow ES}. We report on the potential of student-created artifacts to foster deeper engagement with course material, encouraging both creativity and critical thinking in computing education.
\end{abstract}


\section{Introduction}
Promoting active learning in the Computer Science (CS) classroom has been seen to increase student engagement and foster positive learning experiences \cite{arzaga2023, vasilchenko2020, gallardo2020, forbes2023,vasilchenko2018}. Recent research has increasingly emphasized the importance of metacognitive and communicative skills within technical disciplines such as software engineering. Traditional assessments often emphasize correctness over comprehension, which can lead to surface learning strategies and short-term memorization. Student-created artifacts offer a promising avenue for addressing these needs by engaging students in both the technical and communicative aspects of software design in computing education.

Considering the benefits of active learning in computing education, the challenges observed when instructing higher-level courses in computing, and the current trends of artificial intelligence (AI) in computing education (which use AI tools as passive actors), we considered complementing the students' learning processes by having them design their AI agents for a change. We consider active learning to be optimal when students take action that involves the creation of material that guides them toward learning by teaching. Thus, this Work-in-Progress (WIP) Research-to-Practice Study investigates the efficacy of student-created artifacts as learning tools in a large undergraduate Software Design and Engineering (SWE) course, when the student acts as the designer of multimedia (videos) and intelligent conversational agents to support their learning experiences.

We proposed a learner-centered take-home activity that asked students to deliver student-created artifacts to help a fellow student learn the course content \cite{arzaga2023, vasilchenko2020, gallardo2020, forbes2023, vasilchenko2018}. The assessment had students create two different artifact types: a conversational AI agent (an interactive artifact) and two instructional video tutorials (non-interactive artifacts). We had students from two Summer 2024 courses participate: 1) an introductory course in \textit{Software Design and Engineering}, and 2) a cross-listed senior-level (undergraduate and graduate) course on \textit{User Interface Software Design}. Both courses were offered by the College of Computing of the Georgia Institute of Technology.

We report insights based on students' perceptions of learning and the use of student-created artifacts to support students' learning and motivation in higher-level computing courses.

\section{Background}
The use of student-created artifacts and active learning strategies have been prevalent in the Computer Science Education (CS Ed) community. Active learning has been applied through classroom discussions, online lab environments, group work, learner-generated content \cite{vasilchenko2018, Diemer, Freeman2014, feijoo2020using}, and have also been introduced and evaluated in large computing course formats \cite{feijoo2024learning}. The state of affairs is clear: Active learning strategies offer students across the board better learning opportunities when compared to surface learning strategies \cite{Freeman2014}, with the promotion of structured interactions that bring students from passive to active participants in learning. 

One such way that educators have aided in the active learning push is through student-created artifacts \cite{vasilchenko2018, vasilchenko2020, feijoo2020using, arzaga2023}. Students in this learning strategy create videos, explanation documents, and other materials that help bridge the gap between online and physical environments \cite{arzaga2023}. Past work has even explored the numerical differences in active versus surface learning, finding that students in active learning outperform their surface learning counterparts on exams and course failure rates \cite{Freeman2014}. 

Content creation as well as student-created artifacts have become powerful tools for active learning educators. Past work shows that no matter the STEM corner, these approaches yield numerous benefits for students measured and perceived learning \cite{arzaga2023, gallardo2020, feijoo2020using, feijoo2024learning}. In CS Ed, past work has delved into active learning applications for the transition between abstract and tangible. Student-created artifacts have been shown to work well with physical media, allowing students to visualize component parts that would otherwise be difficult \cite{vasilchenko2020}. To further bridge this gap, peer-to-peer teaching can be employed to yield a wider net of benefit as student-created artifacts are shared and discussed among students \cite{vasilchenko2018, feijoo2020using, feijoo2024learning}.

These aforementioned active learning strategies illustrate the concept of learning-by-design, the sole purpose of which is allowing students to create artifacts that communicate ideas. Past work has agreed that these approaches encourage metacognition and critical thinking. When these effects meet the needs of software engineering and applied computing, the consequences are valuable for both the fields and the students \cite{Reis2022}. When these tools are paired with immediate feedback and collaboration, the use of interactive technologies takes this approach one step further, ensuring an active learning environment \cite{Diemer}.

When looking at the big picture, it's clear that student-generated content, ranging from communicative videos to interactive systems, can support an elevated engagement with course concepts. The implications of this can help students not only reinforce their own learning but also foster a deeper connection to their learning process. 

\section{Intervention}
The strategy was proposed for two high-level Computer Science (CS) courses in summer 2024. We report data from 19 college CS students (out of 43 students in total--44\% of participation) who shared their perceptions discussing how much they learned from designing each student-created artifact. 

The strategy was implemented during the final take-home assessments in both courses. The strategy and data collection were approved by the Institutional Review Board (IRB) of the Georgia Institute of Technology. The first course was an undergraduate introductory course in software design and engineering. The students then responded to four tasks: 

\begin{enumerate}
    \item First Task - Individual doing: The students created a video in which they shared, navigated, and explained a mind map they had to create to explain topics in software design and engineering.
    \item Second Task - Individual doing: The students created a voice-recorded PowerPoint presentation explaining three design patterns. The Factory, Decorator, and Visitor Design Patterns. 
    \item Third Task - Individual doing: The students followed a tutorial provided for them to learn \textit{Google Dialogflow ES} (a tool to build conversational AI agents) \cite{DialogflowDocs}.
    \item Fourth Task - Team-based doing: The students created a conversational AI agent to help explain three software design patterns: Strategy, Composite, and Observer design patterns. 
\end{enumerate}

For the second class, a cross-listed course (i.e., graduate and undergraduate) on principles of user interface software, students responded to three tasks instead:

\begin{enumerate}
    \item First Task - Individual doing: The students read a chapter on Interface Architecture, created a mind map, and created a video where they shared, navigated, and explained the mind map.
    \item Second Task - Individual doing: The students read two chapters. The first on \textit{ accessibility} and the second on \textit{ principles of visual design} in user experience (UX). They then created a voice-recorded PowerPoint presentation in which they described both readings and analyzed the relationship between both topics.
    \item Third Task - Team-based doing: The students were asked to use \textit{Google Dialogflow ES} \cite{DialogflowDocs} to create a conversational agent that helped them explain an academic paper on Human-Computer Interaction with regard to its contribution, research questions, methods, findings, etc.
\end{enumerate}

We asked students to share their perceptions of learning when creating their AI agents and instructional videos, in addition to their insights on what they liked and disliked about creating those artifacts. We used \textit{Qualtrics} to gather student input. Students' \textbf{\textit{perceptions of learning}} were gathered using the questionnaire and method proposed by Diemer et al. \cite{Diemer}. For each type of student-created artifact (video or AI agent), the questionnaire featured six 7-point Likert scales--the word ``\textit{artifact*}" is used in this manuscript to indicate \textit{``video"} or \textit{``conversational agent."}

\begin{enumerate}
    \item ``Creating my \textit{artifact*} helped me apply course content to solve problems."
    \item ``Creating my \textit{artifact*} helped me learn the course content."
    \item ``Creating my \textit{artifact*} helped me me connect ideas in new ways."
    \item ``Creating my \textit{artifact*} helped me participate in the course in ways that enhanced my learning."
    \item ``Creating my \textit{artifact*} helped me develop confidence in the subject area."
    \item ``Creating my \textit{artifact*} helped me develop skills that apply to my academic career and/or professional development."
\end{enumerate}

The students' insights were collected via open-ended questions concerning what they liked and disliked about creating their \textit{artifacts*}:

\begin{enumerate}
    \item ``Share what you \textbf{liked} about creating an \textit{artifact*} to learn course concepts."
    \item ``Share what you \textbf{disliked} about creating an \textit{artifact*} to learn course concepts."
\end{enumerate}

\section{Findings}
Students in both courses reported high levels of perceived learning for both types of artifacts we asked them to create--(\textit{\textbf{Videos}: Avg = 5.43, SD = 1.56}; \textit{\textbf{AI agents}: Avg = 5.59, SD = 1.25}). Moreover, detailing the students' responses to the 7-point Likert scale for the statement,\textit{ "Creating my conversational agent helped me learn the course content,"} we observed that they found creating conversational AI agents to explain their topics helpful to learn course content: \textit{57.9\% (n=11)} students agreed with scores of six and seven.

The students' voices raised interesting insights about their experiences. Concerning the creation of instructional videos, the students appreciated the interactivity the strategy posed:

\begin{itemize}
    \item \textit{``I enjoyed that this gave me a real-world skill."} [P09]
    \item \textit{``The experience felt a lot more interactive than other final exams."} [P03]
    \item \textit{``It's interactive and real-time."} [P01]
\end{itemize}

Also, some of them suggested the strategy to be lengthy and time-consuming:

\begin{itemize}
    \item \textit{``It was a bit lengthy, maybe it should have been released a little earlier."} [P02]
    \item \textit{``Sometimes it feels easier to just present on a Zoom call as opposed to recording a video."} [P03]
\end{itemize}

Regarding the design of their conversational AI agents, students enjoyed the novelty and practical application of the task:

\begin{itemize}
    \item ``I really liked that I learned something completely new that I didn’t even know existed." [P04]
    \item ``Creating the conversational agent definitely gets you thinking like a researcher. It was also gave me practice creating a conversational agent without following the tutorial." [P05]
    \item ``The conversational agent was a unique and interesting approach to apply course content." [P12]
    \item ``I enjoyed learning about conversational agents." [P08]
\end{itemize}

Furthermore, some encountered challenges related to the complexity of using \textit{Google Dialogflow ES} \cite{DialogflowDocs} and the steps required to train their AI agents:

\begin{itemize}
    \item \textit{``Debugging was sometimes a pain as students were not familiar with DialogFlow."} [P03]
    \item \textit{``I disliked training the agent."} [P10]
    \item \textit{``Conversational agents often take you in circles and do not give you great or accurate answers."} [P01]
\end{itemize}

\section{Discussion} \label{Strategy}
Our work contributes insights into the potential of student-created artifacts to foster deeper engagement with course material, encouraging both creativity and critical thinking in computing education. 

Students in both courses reported high levels of perceived learning for both types of
artifacts we asked them to create (Videos: Avg = 5.43, SD = 1.56; AI agents: Avg = 5.59, SD = 1.25). In addition, the students found the creation of AI instructional agents effective in learning course content: 57. 9\% (n = 11) the students agreed with scores of six and seven. The students' voices also raised interesting insights about their experiences. For example, they enjoyed the novelty and practical application of the task concerning the design of their AI instructional agents. 

Overall, the creation of both artifacts (videos and AI agents) brought positive perceptions among students. In particular, the idea of designing conversational AI agents was found to be unique and novel, suggesting that Human-Centered AI can be actively integrated into computing education with students going beyond the use of AI agents by being their designers. 

This study highlights the potential of student-created artifacts to foster engagement, creativity, and critical thinking in computing education. Both instructional strategies were well received, with AI agent design standing out as a unique and promising approach. Our findings suggest that Human-Centered AI can be actively integrated into computing education, shifting students from AI users to AI designers. Future work will explore this strategy at a large scale, navigating the CS curricula from early to senior courses to assess its effectiveness in computing and broader STEM education. 

\section{Limitations}
This pilot study has several limitations that shape both the scope of our conclusions and future research implications. 

Firstly, the study's sample size drew from two small computing courses. This smaller sample size, alongside voluntary participation can limit these findings from being applied to broader undergraduate CS course contexts, especially fall/spring high enrollment seasons. Future iterations will focus on implementing these strategies in large-scale settings, improving the overall student experience and allowing for better statistical reliability.

Secondly, this exploratory study was conducted in the absence of a traditional control group. While students did report high levels of perceived learning, we will strive in future iterations to plot these findings to more traditional end-of-course assessments. Thus, this study may require additional iterations before coming to decisive conclusion regarding student-created artifacts being more or less effective than traditional instructive methods. Future work will hinge on including comparison conditions to assess the relative effectiveness of these interventions.


Thirdly, the instructional strategies in this study utilized overlapping interventions, limiting our ability isolate the direct impacts from each. Measuring comparative effectiveness was not a direct goal of this study, and as such, we are limited in our ability to differentiate factors related to students' reflections on each artifact type.

Fourthly, we acknowledge that this work did not mention certain aspects of an experimental study such as research questions and theoretical frameworks. This work communicates early findings from work that is in progress and still evolving on its methods and approaches. Future iterations will ensure to incorporate more robust experimental components, and include these elements to better communicate research claims and aims.

Finally, though students reported positive responses regarding artifact-based assessments, we are limited in differentiating whether it was the assessments being authentic learning tasks or novel means of assessment that led to these findings. The framing of these interventions in future work will emphasize the pedagogical motivations as well use iterative feedback cycles to further explore student motivations.

\section{Future Directions}
Future research will explore the use of student-created AI in other computing-related formats, considering curricula oriented towards software, hardware, and with aims on different skills: communication, programming, etc. Moreover, we will explore the role of students' situational factors and demographics concerning the effectiveness and acceptance of strategies like this.

Next directions will extend the study toward large course formats in CS Ed, evaluating the impact of the strategy in students' learning outcomes and understanding of course concepts. The experience reported in this manuscript refers to student satisfaction and engagement. Future steps will explore how students' perceived learning relates to their factual learning of course topics, and evaluate the role of strategies like ours in students' professional development as well as their self-efficacy.

\section{Conclusions}
This work-in-progress (WIP) research-to-practice study offers valuable insights into how student-created artifacts can foster deeper engagement with course material, promoting creativity and critical thinking in computing education. Students reported high levels of perceived learning for both instructional videos and AI agents, with particularly strong responses toward the AI design strategy. The creation of these artifacts not only supported their understanding of the content but also introduced elements of novelty and relevance that students found motivating. Particularly, the majority of students rated the AI activity with scores of six and seven, indicating its effectiveness in reinforcing course concepts. Their reflections underscored the value of designing AI agents as a practical and engaging learning experience.

In general, the positive perceptions shared by the students suggest a strong potential to use human-centered AI as a pedagogical strategy. The unique aspect of allowing students to act as AI designers, rather than passive AI users, opens promising directions for enhancing learning in computing disciplines. This shift encourages student agency and a more active role in the creation of educational technologies. Future research may expand the implementation of this strategy across different levels of the computing curriculum, exploring its impact not only in computing, but also in broader STEM education contexts.

\section*{Acknowledgments}

This material is based upon work supported by the National Science Foundation under Grant No. 2434428. Any opinions, findings, and conclusions or recommendations expressed in this material are those of the author(s) and do not necessarily reflect the views of the National Science Foundation.

The authors acknowledge the use of Grammarly and ChatGPT-4o, an AI-based language model developed by OpenAI, for spell checking, grammar, and editing assistance. Both tools were used to refine the language and enhance the manuscript's clarity. No content, figures, or images were generated by AI for this work.

\bibliographystyle{IEEEtran}
\bibliography{references}

\end{document}